# A Computational Framework for Financial Structures

WORKING PAPER

Antonio Scala❏[1], Andrea Monaco❏[2]

❏[1] CNR Institute for Complex Systems, Rome, Italy

❏[2] University College Dublin, Dublin, Ireland

## Abstract

Financial structures transform stochastic cash-flow representations of underlying economic activities into ordered payments across multiple claims through systems of contractual allocation rules. While sophisticated computational models of such structures are widely used in practice, their allocation logic is typically embedded in proprietary implementations and deal-specific documentation, making systematic analysis, comparison, and verification across transactions difficult.

This paper develops a computational framework that formalizes financial structures as structured allocation systems mapping stochastic inflows into hierarchically ordered payments through explicit and state-dependent allocation operators. The framework distinguishes between economic activities, financeable representations, admissible structures, feasible designs, and sustainable designs. It thereby specifies the inputs on which allocation mechanisms operate, the structural requirements that distinguish financial structures from generic contracts, and the feasibility restrictions under which a design may be undertaken.

The contribution of the paper is to show that structurally distinct contractual mechanisms can be represented as allocation operators and evaluated under identical stochastic inflow paths, making differences in expected payments, loss distributions, and sustainability outcomes attributable to contractual design alone.

The framework is not another asset-side cash-flow model. Rather, the paper provides a general representation in which the stochastic generation of inflows is separated from the deterministic and conditional rules governing their distribution across positions. Allocation logic, trigger mechanisms, priority relations, and structural parameters can then be represented within a unified computational architecture.

An illustrative comparison between a sequential waterfall, a reserve-and-trigger waterfall, and a proportional revenue-sharing structure shows how the same stochastic revenue paths generate different payment distributions, loss profiles, and sustainability assessments once passed through different allocation operators.

# 1. Introduction

Financial structures transform uncertain cash inflows generated by underlying assets into ordered payments across multiple claims. Securitisation vehicles, structured credit products, project finance arrangements, and other priority-based contractual frameworks all operate according to this basic logic. In each case, realised inflows are distributed through a system of contractual rules that determine how available funds are applied over time to costs, senior liabilities, and residual claims.

Despite their diversity, these financial structures share a common functional core: each acts as a deterministic allocation mechanism mapping stochastic inflows into hierarchically ordered payments. We refer to such mechanisms as structured allocation systems. They constitute the elementary building blocks of structured finance. While sophisticated computational models of such structures are widely used in industry by rating agencies, financial institutions, and specialised software platforms, the underlying allocation logic is typically embedded in proprietary implementations, deal-specific models, or textual documentation. As a result, the behaviour of financial structures under alternative scenarios is often difficult to analyse, compare, or verify in a transparent and systematic manner across transactions and asset classes.

This lack of an explicit representation is not merely a technical inconvenience. If the allocation logic of a financial structure remains embedded in deal-specific documents or proprietary software, the structure cannot be treated as an object of systematic analysis. Alternative structures cannot be compared under identical stochastic environments, contractual mechanisms cannot be verified independently of their implementation, and structural design cannot be formulated as a transparent computational problem. A general representation of financial structures as explicit allocation operators is therefore needed to make their behaviour inspectable, comparable, and optimisable.

This perspective treats the structure itself as the primary object of analysis, rather than as a fixed environment within which risk is merely measured or priced. It also clarifies why allocation rules, trigger conditions, and structural parameters should be represented explicitly: they determine how identical stochastic inflows are transformed into different distributions of payments across claims.

The modelling of credit risk and structured credit portfolios is well developed (Bluhm and Overbeck 2003; Duffie and Singleton 2003; Andersen and Sidenius 2004; Bluhm et al. 2016). These approaches provide detailed tools for estimating loss distributions, valuing tranches, and managing risk within predefined contractual structures. Industry practice likewise employs advanced waterfall models and scenario-based simulations for transaction analysis. However, both academic and practitioner treatments are typically organised around specific asset classes, transaction types, or proprietary implementations. The computational logic of financial structures is therefore seldom expressed in a unified and explicit form that enables systematic analysis of allocation rules, trigger-driven state transitions, and structural design variables across contexts.

In practice, rating agencies and specialised analytics platforms implement detailed cash-flow models and scenario-based analyses for structured finance transactions, often maintaining extensive libraries of deal-specific waterfall models and associated simulation tools. These implementations provide powerful operational capabilities but are typically proprietary and expressed in procedural or software-specific form, limiting transparency and comparability across transactions.

This paper formalises and generalises the computational logic underlying financial structures by representing them as structured allocation systems acting on stochastic inflow processes. The framework separates the generation of cash inflows from the deterministic and conditional rules governing their allocation across positions, allowing the structure itself to be represented explicitly as a parametric and state-dependent object. Allocation operators, trigger conditions, and priority relations are expressed within a common architecture that makes the behaviour of financial structures analysable, comparable, and, in principle, executable. This separation is the primary methodological innovation of the paper.

By making the allocation logic explicit, the proposed formulation provides a unified conceptual language for describing how contractual design shapes the distribution of risk and return across claims. It supports the computation of scenario-dependent payment streams, valuation measures, and loss distributions. More broadly, representing financial structures as formally specified allocation systems establishes a foundation for the development of executable and verifiable representations of structured finance transactions, in which payment logic and trigger mechanisms can be simulated and inspected as computational objects.

The representation developed here is intentionally general. While motivated by applications in securitisation and structured credit, it applies to any arrangement in which stochastic inflows are allocated across ordered financial claims. The objective is not to replicate existing practitioner tools, but to provide a transparent and unified modelling framework that makes the computational logic of financial structures explicit and amenable to theoretical analysis and systematic design.

The same representation can also describe arrangements outside securitisation when contractual rules allocate stochastic quantities across explicitly defined claims. Two further examples illustrate the scope. In a non-proportional reinsurance treaty, realised losses constitute the stochastic quantity to be allocated: the cedant retains the first layer up to a contractual retention, while the reinsurer absorbs the excess up to a contractual limit. The allocation operator is therefore analogous to a two-position sequential structure. In a revenue-based financing arrangement, a firm allocates realised revenues first to an investor royalty up to a cap, then to operating costs, then to the equity holder; this maps directly onto a three-position sequential structure. The proportional revenue-sharing example in Section 7 adds a third case in which no priority ordering is present. These examples indicate the intended scope of the framework: it applies beyond securitisation when contractual rules allocate stochastic resources across explicitly defined claims. The paper nevertheless focuses on financial structures rather than on allocation mechanisms in general.

The remainder of the paper is organised as follows. Section 2 introduces the conceptual chain from economic activities to financeable representations and then formalises financial structures as allocation operators. Section 3 presents the sequential allocation mechanism and its fundamental computational properties. Section 4 develops a general representation of stochastic inflow generation from heterogeneous cash-flow sources and discrete events. Section 5 introduces a parametric description of positions, trigger mechanisms, and the structural design space. Section 6 discusses performance and risk metrics derived from scenario-dependent payment streams. Section 7 illustrates the comparison of alternative allocation structures under a common stochastic inflow environment. Section 8 relates the framework to structured finance practice and discusses implications for transparency, comparability, and implementation. Section 9 concludes.

## 1.1 Related literature

A large body of work addresses the measurement and modelling of credit risk and structured credit portfolios. Detailed computational approaches to loss distribution estimation, tranche valuation, and portfolio risk management are developed in Bluhm and Overbeck (Bluhm and Overbeck 2003) and Bluhm, Overbeck and Wagner (Bluhm et al. 2016), among others. These frameworks provide sophisticated tools for analysing structured credit positions within specified contractual configurations and stochastic environments.

The economic rationale for structured finance discussed in the literature emphasises risk transfer, tranching, and the design of securities with heterogeneous risk profiles (Boot and Thakor 1993; DeMarzo 2005; Plantin 2004). This strand of research examines why priority-based structures arise and how they allocate risk across investors with different preferences or information. However, the computational representation of the contractual allocation mechanisms themselves typically remains implicit, embedded in deal-specific implementations or pricing models.

From a computational perspective, scenario-based modelling of financial cash flows and credit risk has been widely studied (Duffie and Singleton 2003; Glasserman 2003). In practice, structured finance transactions are evaluated using detailed simulation frameworks in which stochastic asset performance feeds into deterministic payment waterfalls. Yet these implementations are generally transaction-specific and not expressed within a unified formal architecture that isolates the allocation mechanism as an explicit object of analysis.

The present framework contributes by formalising and generalising the computational logic underlying such financial structures. Rather than introducing new asset-side risk models or pricing techniques, it provides an explicit representation of allocation rules, priority relations, and trigger-driven state transitions within a unified computational architecture. In this representation, financial structures become executable allocation systems that can be analysed, compared, and verified independently of the specific stochastic models used to generate inflows.

# 2. From economic activities to allocation operators

## 2.1 Economic activities and financeable representations

Financial structures do not operate directly on economic activities themselves, but on contractual or informational representations of those activities.

At the origin of every financial structure lies an underlying economic process generating value through production, exchange, consumption, or investment. Examples include firms producing goods and services, households making mortgage payments, infrastructure projects generating revenues, or portfolios of receivables producing cash collections over time.

These economic activities become financeable only once they are represented through contractual, legal, or accounting objects that define claims on future cash flows. Loans, bonds, leases, receivables, mortgages, insurance contracts, and similar instruments can therefore be viewed as representations of underlying economic activities that transform uncertain economic outcomes into identifiable financial cash-flow streams.

A financial structure operates on such representations rather than on the underlying activities themselves. Its role is to collect, transform, and redistribute the resulting cash flows across multiple claims according to explicitly specified allocation rules.

This distinction introduces three conceptual layers:

1. an economic layer, consisting of activities that generate value and cash flows;
2. a representation layer, consisting of contractual or legal instruments that define claims on those cash flows;
3. a structural layer, consisting of allocation mechanisms that redistribute realised cash flows across financial positions.

The framework developed in this paper focuses primarily on the third layer. However, making the distinction explicit clarifies the scope of the model and highlights its generality. Different economic activities may generate identical financial representations, and identical financial representations may be embedded within different allocation structures. Separating these layers therefore facilitates comparison, analysis, and design of financial structures independently of the specific nature of the underlying assets.

Within this perspective, financial structures can be viewed as computational mechanisms acting on financeable representations of economic activity, transforming stochastic inflows into ordered distributions of payments across claims.

## 2.2 Admissibility requirements for financial structures

Before specifying the allocation operator, we make explicit the conditions an arrangement must satisfy to constitute a *financial structure* in the sense of this paper, as opposed to a generic contractual arrangement. These conditions are stated as requirements on the inputs of the structure – the financeable representations and their stochastic model – and on the objects it produces – the positions. They are definitional and structural: no pricing, preference, or market-equilibrium model is introduced. Their role is to delimit the class of admissible structures on which the allocation operator acts, and thereby to separate financial structuring from arbitrary contracting. The requirements are deliberately general; they apply equally to non-marketable arrangements, such as bilateral or project-specific structures, and to marketable ones whose positions are securities.

We group the requirements by the layer to which they apply.

**Requirements on the financeable representations (asset side).**

(R1) *Cash-flow generation.* Each underlying representation $i$ produces a well-defined, non-negative cash-flow process, and the aggregate inflow process $\{F_t(\omega)\}_{t=0}^{T}$ with $F_t(\omega) \geq 0$ is the object on which the structure operates.

(R2) *Recognised value.* Each representation carries an identifiable, contractually or legally recognised claim on those future cash flows – that is, it admits a value capable of serving as collateral for the structure. The framework does not compute this value; it requires that such a value exist and be identifiable. A mere expectation of cash unaccompanied by a recognised claim is not a financeable representation.

(R3) *Identifiable risk factor.* The loss-generating mechanism is identifiable: there exists a specified event or state process $\sigma_{i,t}(\omega)$ (Section 4) whose realisations determine the deviations of realised cash flows from their baseline $b_{i,t}$. Losses must be attributable to identified factors rather than to an unmodelled residual.

(R4) *Specified dependence.* When the pool comprises more than one representation, the joint law of the unit processes – their dependence structure – is part of the model specification. Pooling without a specified dependence structure does not define an admissible inflow process.

**Requirement on the positions (liability side).**

(R5) *Inspectable allocation.* The allocation operator is finitely and explicitly specified, so that the mapping from realised inflows and state to position payments is inspectable by the holder of any position. Inspectability is the structural counterpart of the requirement that a holder be able to locate the risk borne by a position. In securitisation markets, this requirement also has a regulatory analogue: Article 22(1)(c) of Regulation (EU) 2017/2402 requires that investors be given access to a liability cash-flow model before pricing, and to updated liability cash-flow information on an ongoing basis. The present framework abstracts this requirement into a general structural condition: the allocation logic must be explicit enough to be inspected, simulated, and verified independently of any proprietary implementation. This legal-technical interpretation is developed in greater detail in recent work on executable waterfall representations (Scala 2025).

**Definition (admissible financial structure).** A financial structure is *admissible* if its inflow process satisfies (R1)–(R4) and its allocation operator satisfies (R5). The allocation operator introduced below acts on admissible structures.

These requirements make precise the distinction between a financial structure and a generic contractual arrangement. The distinction rests on the combination of identifiable, value-bearing claims (R1–R2), an identifiable loss-generating mechanism and, for pooled inflows, a specified dependence structure (R3–R4), and an inspectable allocation of those inflows across multiple ordered positions (R5 together with the priority structure of the operator). An arrangement lacking these features is not a financial structure in the present sense, even if it moves cash between parties. Transferability of the claims is not among the admissibility requirements: the framework covers both non-marketable structures – bilateral or project-specific arrangements whose claims are not assignable – and marketable structures whose positions can be transferred independently of their holders.

**Classification (marketable structures).** When, in addition to (R1)–(R5), the positions of an admissible structure are transferable independently of the identity of their holders, the structure is *marketable* and its positions are securities.

Marketability is therefore a classifying property identifying a subclass of admissible structures, not a condition for admissibility, and the existence of a market on which transfer occurs is treated as exogenous. The same allocation operator and the same analysis apply across the subclass and its complement; securitised structures are the marketable special case of the general object studied here.

## 2.3 Minimal primitives

We consider a financial structure that transforms uncertain cash inflows into payments toward a set of ordered claims. Such structures arise in securitisation, project finance, structured credit, and other settings in which stochastic resource streams must be allocated across hierarchical liabilities.

Let time be discrete $t=0,1,\ldots,T$.

Let $F_t(\omega)$ denote the gross cash inflow at time $t$ under scenario $\omega$ in a probability space $(\Omega,F,P)$. The process $F_t$ represents the aggregate inflows generated by an underlying pool of exposures or assets. No assumption is required at this stage regarding the origin or statistical structure of $F_t$, except non-negativity:

$$F_t(\omega)\geq 0\;\forall t,\omega.$$

Let the liability side consist of $P\geq 1$ positions indexed by $p=1,\ldots,P$. Each position represents a contractual claim on the available cash flows. Positions are ordered by a strict priority relation $\prec$ such that if $p\prec q$, then position $p$ has higher payment priority than position $q$.

Let $P_{p,t}(\omega)$ denote the payment made to position $p$ at time $t$ under scenario $\omega$.

A financial structure is characterised not only by the realised inflows but also by a collection of state variables describing the contractual configuration of the structure at time $t$.
Let $S_t(\omega)$ denote the structural state at time $t$, which may include outstanding tranche balances, reserve accounts, trigger statuses, cumulative losses, and any other variables required by the contractual rules.

The evolution of the structure is then described by a deterministic allocation operator

$$(P_{1:P,t}(\omega),S_{t+1}(\omega))=A_t\left(F_t(\omega),S_t(\omega)\right),$$

which maps current inflows and the current structural state into a vector of payments across positions and an updated structural state for the next period.

Equivalently, the operator may be written as a mapping from the inflow history and current state,

$$A : \left( \{ F_\tau \}_{\tau \le t}, S_t \right) \mapsto \left( \{ P_{p,t} \}_{p=1}^{P}, S_{t+1} \right),$$

but in most practical implementations the Markovian representation above is sufficient, since all relevant historical information is encoded in the state vector $S_t$.

The operator $A_t$ encodes the contractual payment rules of the structure, including priority of payments, admissibility conditions, balance updates, reserve dynamics, and any state-dependent constraints on distributions. By construction, the operator is deterministic: given realised inflows and the current structural state, both payments and the next structural state are uniquely determined.

Determinism is understood conditional on realised inflows and state. Apparent randomness in payment outcomes arises exclusively from stochastic inflows or from event-driven transitions of the structural state (for example, defaults that reduce outstanding balances or trigger activations that modify allocation rules). Such events alter the state variables or contractual parameters governing the operator, but the mapping from realised inflows and state to payments and next state remains deterministic. The framework therefore represents financial structures as state-dependent deterministic allocation mechanisms operating on stochastic inputs.

This representation isolates the fundamental object of interest: a financial structure is a deterministic state machine driven by stochastic inflows. Figure 1 summarises the computational architecture of the framework, from economic activities and financeable representations to stochastic inflows, deterministic allocation, position-level outputs, and structural analysis.

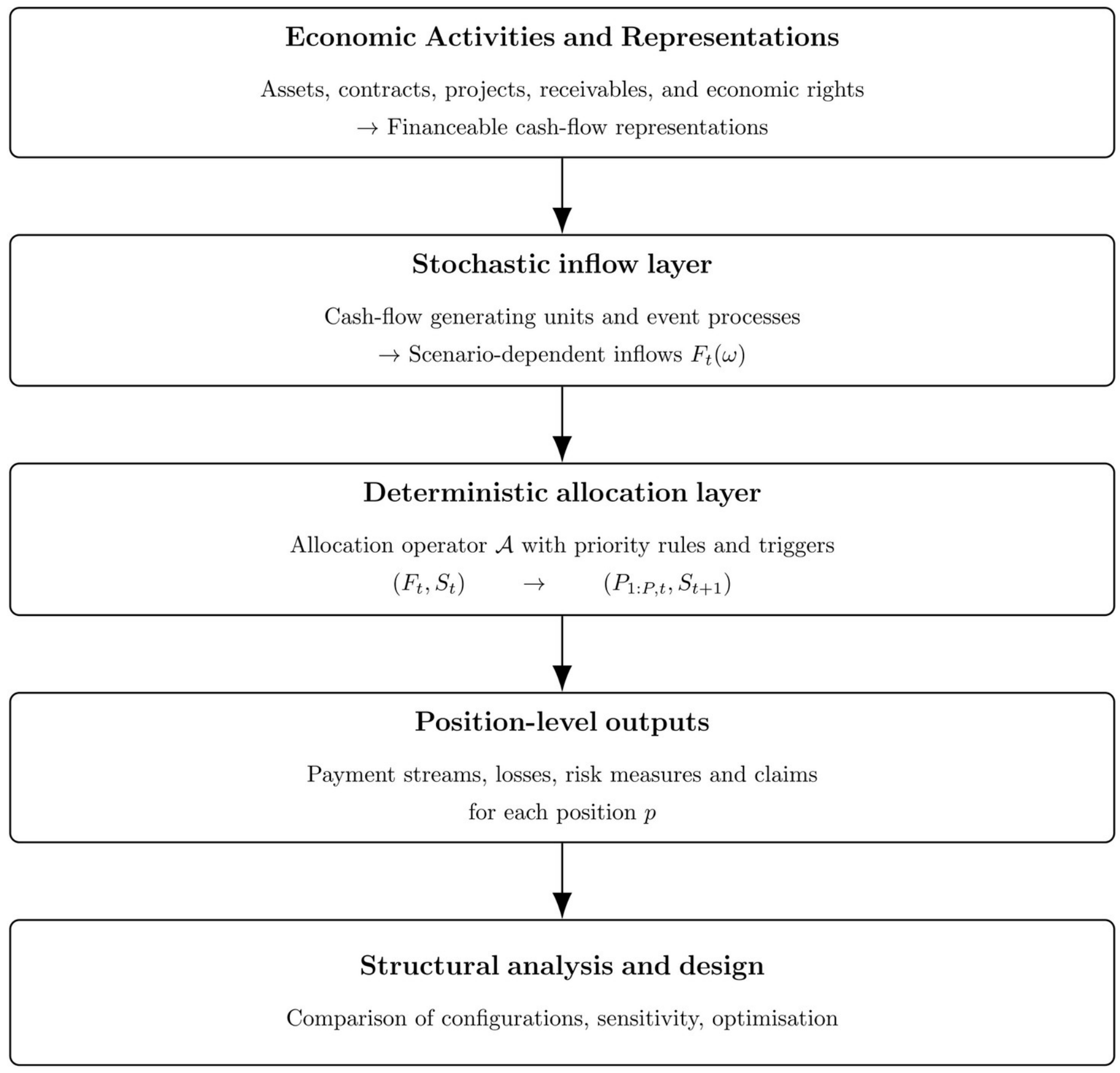

*Computational architecture of the framework. Economic activities generate financeable representations, which produce stochastic inflows. These inflows feed a deterministic allocation operator, producing position-level payment streams used for structural analysis and design.*

## 2.4 Allocation constraints

Any admissible allocation operator must satisfy basic consistency properties.

**Non-negativity.** Payments cannot be negative:

$$P_{p,t}(\omega) \geq 0 \; \forall \, p, t, \omega.$$

**Conservation of funds.** Total payments cannot exceed available resources. Let $R_t(\omega)$ denote liquid funds carried into period $t$ (for example, cash accounts or reserves), included within the structural state $S_t(\omega)$. Then

$$\sum_{p=1}^{P} P_{p,t}(\omega)+R_{t+1}(\omega)=F_t(\omega)+R_t(\omega),$$

with $R_t(\omega)\geq 0$ for all $t$.
The residual $R_{t+1}$ forms part of the next structural state $S_{t+1}$. In the general framework, $R_t$ denotes the liquid-resource component of the structural state. Depending on the application, it may represent residual cash, a reserve account, or another explicitly modelled liquidity account. When condition (V4) is imposed below, it applies to the particular liquid-resource component selected as relevant for operational sustainability.

**Priority consistency.** If $p<q$, then payments to $q$ at time $t$ can occur only after all admissible payments to $p$ at time $t$ have been satisfied according to the contractual rules encoded in $A$.

**Resource monotonicity.** For a fixed allocation regime, larger realised inflows weakly increase the aggregate resources available to the structure. At the position level, however, monotonicity need not hold when trigger mechanisms, reserve accounts, or regime-switching allocation rules are present. Additional inflows may be diverted to reserve replenishment, deleveraging, or other contractual objectives before reaching a given position. Position-level monotonicity is therefore a property of fixed-regime sequential waterfalls, not of all admissible allocation operators.

These properties define a class of admissible allocation operators suitable for modelling financial structures.

## 2.5 Sequential allocation and waterfall structures

In most practical applications, the operator $A$ takes the form of a sequential allocation rule. Available funds at time $t$ are applied in a predetermined order: first to senior costs or positions, then to progressively more junior claims. Only once higher-priority obligations are satisfied can residual funds be allocated further down the hierarchy or retained.

Let $C_{p,t}(S_t)$ denote the contractual amount due to position $p$ at time $t$ (interest, principal, or cost), which may depend on the current structural state through outstanding balances, accrued amounts, or trigger conditions. Let $A_t=F_t+R_t$ denote total available funds at time $t$; a generic sequential allocation can then be written recursively as

$$P_{p,t}=min\left(C_{p,t}, A_t-\sum_{q<p} P_{q,t}\right),$$

with residual

$$R_{t+1} = A_t - \sum_{p=1}^{P} P_{p,t}.$$

This formulation makes explicit that the allocation at each level is determined uniquely by remaining available funds and contractual parameters. The structure is therefore fully determined by the inflow process, the contractual obligations $C_{p,t}$, and the priority ordering.

## 2.6 Scope of the framework

The representation above is intentionally minimal. It does not impose any specific structure on the inflow process, the nature of contractual obligations, or the origin of positions. Instead, it defines a general computational architecture for any system in which stochastic inflows must be allocated across hierarchically ordered claims.

In software terms, the allocation operator can be implemented as a pure function: given an inflow path, an initial state, and a set of contractual parameters, it returns payment paths and state trajectories without altering the stochastic input layer. This makes the representation portable across implementation environments such as Python, R, Julia, or domain-specific transaction-modelling systems.

In subsequent sections we introduce a general stochastic framework for generating inflows from heterogeneous cash-flow sources subject to discrete events, together with a design space for the construction of positions and the evaluation of their risk and performance. Securitisation and structured credit motivate several examples, but the comparative illustration uses a generic revenue environment to emphasise the generality of the allocation-operator representation.

## 2.7 Economic interpretation

Although introduced as a computational representation, the allocation operator also has a direct economic interpretation. A financial structure can be viewed as a contractual mechanism that distributes uncertain cash flows across agents holding heterogeneous claims. Priority ordering, position size, and trigger conditions together determine how gains and losses are allocated across participants under different realisations of the stochastic environment.

Sequential allocation rules implement a specific loss-sharing mechanism. Senior claims receive priority access to available funds and therefore bear lower risk, while junior positions absorb initial losses and provide credit enhancement. Residual or equity positions absorb remaining variability once contractual obligations have been satisfied. In this sense, the structure defines an ex-ante distribution of risk and return across participants, implemented mechanically through deterministic allocation rules.

Trigger conditions and state-dependent allocation rules introduce regime changes in this distribution. When structural thresholds are crossed – for example, through performance deterioration or depletion of reserves – payment priorities or allocation fractions may change, redistributing risk across positions. These mechanisms can be interpreted as contractual responses to realised states of the world, designed to preserve senior protection or maintain structural stability.

By representing these features explicitly within an allocation operator, the framework makes the economic logic of the structure transparent. Contractual design becomes analysable as a mapping from stochastic environments to distributions of payments across claims.

## Deterministic contractual allocation and renegotiation

The framework represents financial structures as deterministic allocation systems conditional on realised inflows and formally specified state variables. This corresponds to the operational logic of many structured finance and priority-based contractual arrangements, in which payment rules and trigger conditions are explicitly defined and mechanically enforceable.

In practice, some financial arrangements may be subject to renegotiation, discretionary management actions, or informal interventions that alter contractual payment patterns. Such features lie outside the baseline representation developed here. The objective of the framework is to formalise and analyse rule-based allocation mechanisms whose behaviour is fully determined by contractual parameters and realised states.

When discretionary or renegotiated outcomes can be represented through explicit policy rules or state transitions, they may be incorporated within the same computational architecture by extending the state space or introducing additional allocation regimes. The baseline formulation nevertheless focuses on structures whose allocation logic can be expressed as a deterministic mapping from stochastic inflows and state variables to payments across positions.

# 3. Sequential allocation and computational implementation

## 3.1 The allocation algorithm

We now make the allocation operator explicit. At each time $t$, available funds are allocated across positions according to a deterministic priority rule. By defining * $F_t(\omega)$ to be the realised inflow at time $t$ * $R_t(\omega)$ the residual funds carried from the previous period * $A_t(\omega)=F_t(\omega)+R_t(\omega)$ the total available funds * $C_{p,t}(\omega)$ the contractual amount due to position $p$ at time $t$ * the positions indexed so that $p=1$ is most senior and $p=P$ most junior the allocation operator can be implemented as a recursive state-transition rule operating on current inflows and the residual state carried from the previous period.

Thus, for each period $t$ and scenario $\omega$: 1. set remaining funds

$$A \leftarrow A_t(\omega)$$

2. for $p=1,\ldots,P$ (in priority order): * compute admissible payment

$$P_{p,t}(\omega)=min\left(C_{p,t}(\omega),A\right)$$

* update remaining funds

$$A \leftarrow A-P_{p,t}(\omega)$$

3. carry forward the residual:

$$R_{t+1}(\omega)=A$$

This sequential rule defines a deterministic mapping from realised inflows to payments. Given $(F_t,R_t)$ and contractual obligations $C_{p,t}$, the resulting payment vector is uniquely determined.

## 3.2 Fundamental properties

The sequential allocation rule satisfies the following properties:

**Conservation of funds**

$$\sum_{p=1}^{P} P_{p,t}(\omega)+R_{t+1}(\omega)=F_t(\omega)+R_t(\omega).$$

**Priority consistency** If $p<q$, then payments to position $q$ occur only after admissible payments to all positions $1,\ldots,p$ have been satisfied.

**Monotonicity within the basic sequential rule** Let $\widetilde{F}_t(\omega)\geq F_t(\omega)$ for all $t$. Then, within a fixed sequential regime, payments to more senior positions cannot decrease:

$$\widetilde{P}_{p,t}(\omega)\geq P_{p,t}(\omega)$$

for all $p$ until contractual caps become binding.

These properties hold for the basic sequential allocation rule independently of the origin of inflows. Trigger-driven or reserve-dependent regimes may require the caveat stated in Section 2.4 and Appendix A.

## 3.3 Computational implementability of the allocation operator

The sequential allocation rule above defines a finite deterministic state–transition algorithm. Given realised inflows and the current residual state, the payment vector and the updated residual are uniquely determined through a linear recursive procedure across positions.

For each scenario and period, evaluation requires a single pass across the ordered set of positions. The computational cost per period is therefore linear in the number of positions $P$, and the total evaluation cost per scenario over a horizon $T$ is $O(PT)$. No fixed-point iteration or optimisation step is required within the allocation stage; the operator is directly executable.

Scenario evaluations are independent and can be parallelised. Moreover, the stochastic inflow process can be simulated once and reused across multiple contractual configurations or parameter choices. This separation between inflow generation and allocation allows alternative structural specifications to be evaluated under identical stochastic environments, ensuring strict comparability of outcomes.

Because the operator is deterministic conditional on realised inflows and structural state, the full distribution of position-level payments is obtained by applying the same allocation rule across a collection of simulated inflow paths.

This computational structure makes financial contracts amenable to systematic analysis and design. Structural parameters such as position sizes, priority orderings, trigger thresholds, or reserve policies can be varied and evaluated within a common simulation environment. The framework thus supports sensitivity analysis, structural comparison, and constrained optimisation over contractual configurations while preserving a transparent mapping from realised inflows to payments.

Subsequent sections develop a stochastic environment for inflow generation and use the executable allocation operator to study how alternative structural parameterisations affect the distribution of payments and losses across positions.

# 4. Stochastic generation of cash inflows

## 4.1 Cash-flow generating units

We now construct the stochastic inflow process introduced in the previous sections. Financial structures receive cash flows generated by underlying economic or financial units such as loans, leases, receivables, bonds, or real assets. We refer to these generically as cash-flow generating units. The inflow-generating units may equivalently be interpreted as independent stochastic processes – or, in a broader computational economics perspective, as elementary agents – whose state-dependent cash flows contribute to the aggregate inflow process. This interpretation facilitates integration of the framework within larger simulation environments in which multiple stochastic components interact.

Let units be indexed by $i=1,\dots,N$. Each unit produces a time-indexed cash-flow process whose realised value depends on contractual terms, economic conditions, and discrete events.

For each unit $i$, we distinguish between a baseline cash-flow component and a set of state-dependent modifications. The baseline component represents scheduled or expected payments under normal performance. Depending on asset type, this component may be deterministic, stochastic, or null. For example:

- amortising loans generate scheduled principal and interest payments;
- revolving receivables generate stochastic but persistent inflows;
- project-finance assets produce cash flows linked to output or prices;
- certain exposures may generate flows only conditional on specific events.

Let $b_{i,t}$ denote the baseline cash flow of unit $i$ at time $t$. This baseline may itself be deterministic or stochastic.

## 4.2 Event-driven dynamics

Realised cash flows typically deviate from baseline schedules due to discrete events. Examples include default, prepayment, restructuring, recovery, or other contractual or economic contingencies. These events alter the timing or magnitude of payments and may terminate or transform future flows.

Let $E_i$ denote the set of relevant event types for unit $i$. Each unit is associated with an event process determining the occurrence and timing of events in $E_i$.

We represent the state of unit $i$ at time $t$ by a state variable $\sigma_{i,t}(\omega)$ capturing the cumulative effect of events up to time $t$ in scenario $\omega$. The realised cash flow of unit $i$ at time $t$ is then given by a transformation

$$F_{i,t}(\omega) = \Phi_i\left(b_{i,t}, \sigma_{i,0:t}(\omega)\right),$$

where $\Phi_i$ encodes contractual rules governing how events modify baseline payments. For example:

- default may terminate scheduled payments and trigger recovery;
- prepayment may replace future scheduled flows with a lump-sum payment;
- restructuring may alter future payment levels or timing.

This representation accommodates both schedule-based and fully stochastic assets. If a unit has no predetermined schedule, the baseline component $b_{i,t}$ may be interpreted as a stochastic production process, with events representing regime changes or interruptions.

## 4.3 Scenario construction

A scenario $\omega$ specifies a realisation of event processes across all units and all times. Given a scenario, the state trajectories $\sigma_{i,t}(\omega)$ determine the realised cash flows $F_{i,t}(\omega)$ for all units.

The joint scenario space is generally high-dimensional. Even when individual units have a small number of possible events, the combination across units and time yields a combinatorial growth in possible trajectories. For practical analysis, the inflow process is therefore evaluated using sampling or simulation methods that generate representative scenarios from the underlying event dynamics.

The modelling framework does not require a specific probabilistic structure for event processes. Correlations, common factors, or dependence across units can be incorporated as needed. What matters for the allocation problem is that each scenario yields a well-defined path of realised inflows.

## 4.4 Aggregation of inflows

Total inflow to the financial structure at time $t$ is obtained by aggregating realised flows across all units:

$$F_t(\omega) = \sum_{i=1}^{N} F_{i,t}(\omega).$$

The process $F_t(\omega)_{t=0}^{T}$ constitutes the stochastic inflow sequence entering the structured allocation system defined in Section 2. Combined with the deterministic allocation operator, it fully determines the distribution of payments across positions under each scenario.

## 4.5 Relation to existing securitisation modelling approaches

The event-based construction described above provides a general representation of how stochastic inflows arise in structured finance. In the specific context of securitisation, classical treatments of credit portfolio and tranche modelling provide detailed methodologies for asset-level cash flows, credit events, loss distributions, and tranche risk measures; see, for example, Bluhm and Overbeck (2003) and Bluhm, Overbeck and Wagner (2016). More recent securitisation-specific attempts, such as the PEAL approach (Pinto and Scala 2024), have proposed modular representations linking asset-side scenarios, position design, waterfall configurations, and risk features. The present framework abstracts from these securitisation-specific constructions and focuses on a more general problem: representing heterogeneous financial structures as allocation operators that can be compared under common stochastic inflow paths.

Beyond providing a unified representation for analysis, the framework developed here establishes a computational basis for the explicit encoding and execution of financial structures as algorithmic systems. When allocation rules, trigger conditions, and state variables are represented within a common formal architecture, structured finance transactions can be specified as executable objects rather than purely descriptive contractual arrangements.

Recent work has begun to explore this direction by proposing executable representations of payment waterfalls and trigger-driven allocation logic within securitisation structures (Scala 2025). The framework introduced in this paper provides the general modelling foundation required for such developments by representing financial structures as state-dependent deterministic allocation systems acting on stochastic inflows.

# 5. Structural design of financial positions

## 5.1 Positions, state, and conditional priority

A financial structure allocates realised inflows across a set of contractual claims (positions). In many transactions the allocation rules are conditional: trigger tests (e.g. over-collateralisation, interest coverage, cumulative loss ratios, revolving termination, clean-up calls) may redirect cash flows, defer principal, or switch the waterfall family once specified conditions are met.

Accordingly, positions are evaluated relative to a structure-level state variable $S_t(\omega)$, which encodes tranche balances, reserve accounts, trigger statuses, and other contractual context. Allocation rules are then functions of both available funds and state.

## 5.2 Parametric representation including triggers

For each position $p$, define a contractual profile

$$\Theta_p = (\acute{N}_p, \pi_p, \tau_p, \Gamma_p),$$

where $\Gamma_p$ now includes any position-specific contractual parameters (rates, caps, step-ups) and may reference structure-level trigger variables.

In addition, define a structure-level parameter set $\Gamma$ collecting trigger definitions and thresholds. Trigger tests are represented as Boolean functions of the current state, e.g.

$$\chi_j(S_t(\omega), t) \in \{0, 1\},$$

and allocation rules are permitted to depend on $\{\chi_j\}$, yielding both pure and conditional waterfalls within the same operator architecture.

## 5.3 Structural configurations

Financial structures differ not only in the number of positions but also in how risk and cash flows are distributed across them. Two broad configuration principles can be distinguished.

**Sequential (vertical) allocation.** Positions are ordered strictly by priority. Losses and shortfalls are absorbed first by the most junior positions and propagate upward only after junior capacity is exhausted.

**Proportional (horizontal) allocation.** Available funds or losses are distributed across selected positions according to predefined proportions. Such arrangements may arise in co-investment structures, pari passu tranches, or hybrid allocation rules.

**Hybrid configurations.** Most real-world financial structures combine sequential and proportional elements. Positions may be grouped into tiers within which payments are shared proportionally, while tiers themselves are ordered sequentially.

The allocation operator introduced earlier accommodates all such configurations through appropriate specification of priority relations and payment rules.

## 5.4 Design degrees of freedom

Once positions are represented parametrically, the structure becomes a design object. Key design variables include:

- number and ordering of positions;
- thickness or notional of each position;
- maturity profiles;
- allocation rules within and across tiers;
- presence of reserve or trigger mechanisms.

These choices determine how variability in inflows translates into variability of payments across positions. In particular, they govern the distribution of expected returns, risk exposure, and loss absorption across the structure.

The framework developed here treats these structural characteristics as explicit parameters rather than fixed institutional features. This allows different configurations to be compared and analysed within a unified computational setting.

## 5.5 From allocation to optimisation

Given a stochastic inflow process and a parametric description of positions, the allocation operator produces scenario-dependent payment streams for each position. This enables the computation of performance and risk measures such as expected payments, shortfall probabilities, and loss distributions under alternative structural configurations.

Structural design may therefore be formulated as an optimisation problem: selecting position parameters, trigger conditions, and priority relations to achieve specified objectives under uncertainty. Objectives may include return targets, risk constraints, capital or regulatory requirements, or other performance criteria. Optimisation-based approaches to tranche design and asset selection have been developed within the securitisation literature, typically in deal-specific contexts and for particular asset classes (Mansini and Speranza 1999). Within the present framework, the emphasis is not on introducing new optimisation techniques, but on providing a unified computational representation in which alternative structural configurations and design objectives can be formulated and evaluated consistently across contexts.

The separation between inflow generation and allocation established in the previous sections allows alternative structural designs to be analysed within a common stochastic environment. This provides a systematic basis for comparing configurations and for exploring design trade-offs while preserving a transparent representation of the allocation logic governing payments across positions.

## 5.6 Economic viability and the feasible design set

The requirements of Section 2.2 admit a structure as well-formed; whether a *given design* is one that would actually be undertaken is a separate question. We capture it as a feasibility restriction on the design space, expressed entirely through quantities the operator already produces and through thresholds supplied exogenously, without introducing a pricing or funding model.

Let a design be described by the position profiles $\Theta=\{\Theta_p\}$ and the structure-level parameters $\Gamma$ of this section. We use the following position-level quantities, defined formally in Section 6 below: cumulative payments $\Pi_p(\omega)$, cumulative losses $L_p(\omega)$, the impairment probability $\psi_p(\Theta)$, and the expected payment multiple $m_p(\Theta)$. A design is *economically feasible* if it simultaneously satisfies the following conditions.

(V1) *Originator participation.* Let $O \subseteq \{1,\ldots,P\}$ denote the set of positions retained by the party undertaking the structure (the originator or sponsor). In many financial structures, regulatory or contractual provisions require the originator to retain a non-empty economic interest, $O \neq \varnothing$. When such requirements apply, originator participation is assessed through the retained set $O$. The originator bears a structuring and regulatory cost $\kappa \geq 0$ and has an exogenous outside option $\rho_0$, for instance the return obtainable by funding the activity by other means. When a retained position exists, the design is originator-feasible if

$$\sum_{p \in O} E\left[\Pi_p(\Theta)\right] - \kappa \geq \rho_0 .$$

The retained set $O$ may consist of the residual position, of a single position, or of several; in each case originator participation is assessed entirely from the cumulative payments the operator already produces, with no inception valuation. Structures in which the originator places every position remain admissible and fully analysable under Sections 2–6; their originator-side viability would depend on the proceeds raised at inception and is therefore a pricing question, lying outside the feasibility analysis developed here.

(V2) *Investor participation.* Each placed position $p$ (that is, $p \notin O$) must meet the acceptability threshold of its intended holder, expressed as a restriction on the position's payment or loss distribution,

$$\psi_p(\Theta) \leq \varepsilon_p \,\text{and/or}\, m_p(\Theta) \geq \tau_p ,$$

for exogenous tolerances $\varepsilon_p$ (maximum admissible impairment probability) and $\tau_p$ (minimum admissible expected multiple). More generally, acceptability may be written as $u_p\left(\Pi_p(\Theta)\right) \geq \acute{u}_p$ for an exogenous acceptability functional $u_p$ acting on the distribution of $\Pi_p$; the threshold form above is the operational special case used here.

(V3) *Regulatory feasibility.* Binding regulatory conditions – capital, retention, or eligibility requirements – enter both as additional restrictions on $\Theta$ and through the cost term $\kappa$ in (V1).

(V4) *Liquidity sustainability.* Let $R_t(\omega)$ denote the liquid-resource component of the structural state selected for sustainability assessment. This component may coincide with residual cash, with a reserve account, or with another explicitly modelled liquidity account. A design is liquidity-sustainable if the selected liquid-resource component remains above a prescribed buffer level $B \geq 0$ with sufficient probability:

$$Pr\left(\min_{t \leq T} R_t(\omega) < B\right) \leq \eta .$$

The parameter $B$ represents the minimum liquidity buffer required for the structure to remain operational, while $\eta$ is the tolerated probability of breaching this buffer. In practical implementations, breach of the buffer may trigger reserve replenishment, liquidity facilities, equity support, or other contractual support mechanisms. Such mechanisms can be represented within the framework as additional state variables, positions, or trigger-dependent allocation rules.[1]

The conditions (V1)–(V3) define economic feasibility, while (V4) captures operational sustainability.

---

[1]

The probability-based sustainability condition can be strengthened, for example by introducing a severity constraint. Let

$$\Delta = E ¿$$

This quantity measures the expected liquidity support required to restore the structure above the minimum buffer level. In settings where sponsors or equity holders provide committed liquidity facilities, sustainability may therefore be formulated jointly in terms of breach probability and expected support capacity. The present paper adopts the simpler probability-based condition for expositional clarity.

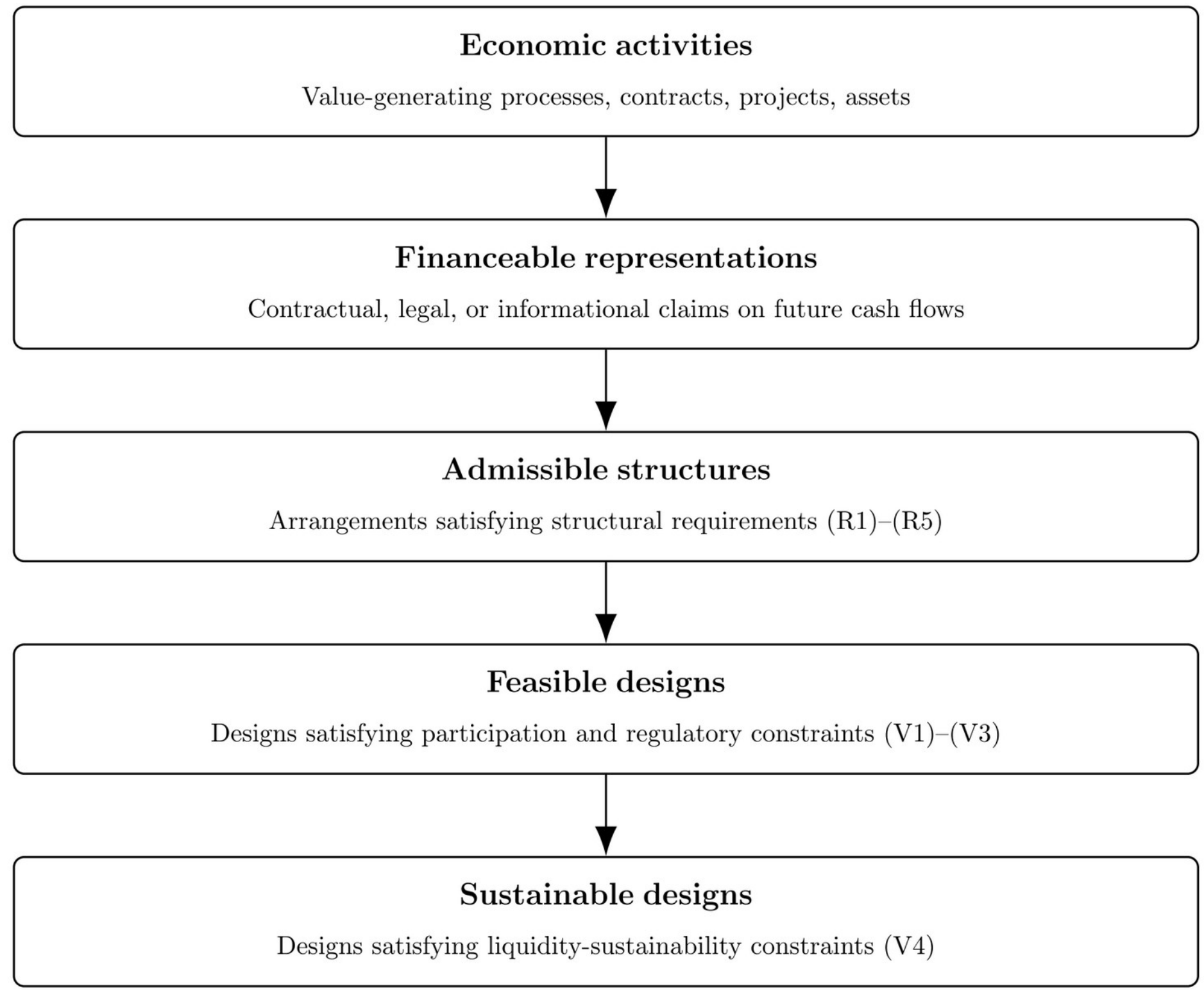

*Conceptual hierarchy of the framework. Economic activities are first represented as financeable cash-flow objects; admissibility requirements identify financial structures; feasibility and sustainability constraints then restrict the set of designs that can be undertaken and remain operational.*

The feasible design set is then

$$Feasible = \left\{ \Theta : \backslash(\mathrm{V1}\backslash),\ \backslash(\mathrm{V2}\backslash),\ \backslash(\mathrm{V3}\backslash),\ \backslash(\mathrm{V4}\backslash)\ \text{hold} \right\},$$

and the structural design problem of this section is posed over it,

$$\max_{\Theta \in Feasible} J\left(\left\{ P_{p,t}(\omega) \right\}\right).$$

A design that satisfies (R1)–(R5) but violates one or more of (V1)–(V4) is structurally admissible but not operationally viable. It is therefore an exercise in allocation rather than a financial structure in operation. Together, the admissibility requirements and the feasibility set delimit financial structures from both directions: by the conditions an arrangement must satisfy to qualify, and by the conditions a design must satisfy to be undertaken and remain sustainable.

These conditions keep the framework within computational economics. The structure is characterised, and its admissible designs are delimited, entirely by conditions on computable cash-flow quantities and on supplied behavioural thresholds, without recourse to a pricing model or to market-equilibrium assumptions. The economic content enters as constraints on the mechanism, not as the mechanism itself.

### 5.7 Computational tractability

From a computational standpoint, the allocation stage is lightweight relative to scenario generation. For each period, sequential allocation requires at most a linear scan across positions, implying complexity proportional to the number of positions per period. By contrast, the generation of stochastic inflow scenarios typically dominates computational cost, especially when large numbers of underlying units or correlated event processes are modelled.

The explicit separation between stochastic inflow generation and deterministic allocation enables modular implementation. Inflow scenarios can be generated once and reused across multiple structural configurations, allowing systematic comparison without repeated simulation of the underlying asset layer. Because allocation across scenarios is independent conditional on the inflow paths, evaluation can also be parallelised efficiently across scenarios or structures. This modular architecture makes large-scale structural evaluation and sensitivity analysis computationally tractable even when the underlying stochastic environment is high-dimensional.

## 6. Performance and risk metrics

Performance and risk metrics are computed from the scenario-dependent payment streams generated by the allocation operator. Once structures are expressed within a common computational representation, different contractual designs can be evaluated under the same inflow scenarios and performance criteria.

### 6.1 Scenario-dependent payment streams

Given a stochastic inflow process $F_t(\omega)$ and a specified set of positions $\Theta_p$, the allocation operator determines realised payments

$$P_{p,t}(\omega)$$

for each position $p$, time $t$, and scenario $\omega$.

These payment streams constitute the primary output of the structured allocation system. All performance and risk metrics associated with the structure can be derived from them. The framework therefore reduces the evaluation of financial structures to the analysis of scenario-dependent payment vectors generated by a deterministic allocation rule acting on stochastic inflows.

## 6.2 Expected payments and valuation

The expected payment profile of position $p$ is given by

$$E[P_{p,t}] = \sum_{\omega \in \Omega} P_{p,t}(\omega) P(\omega),$$

or by its simulation analogue under sampled scenarios.

From these expected payments, standard valuation measures can be constructed. For example, the present value of position $p$ under a discount curve $d_t$ is

$$V_p = \sum_{t=0}^{T} d_t E[P_{p,t}].$$

This representation emphasises that valuation depends jointly on the stochastic generation of inflows and the structural allocation rules governing payments.

## 6.3 Loss and shortfall measures

Risk associated with a position can be expressed in terms of deviations of realised payments from contractual or target levels.

Let $C_{p,t}$ denote the contractual due for position $p$ at time $t$. Define cumulative loss under scenario $\omega$:

$$L_p(\omega) = \sum_{t=0}^{T} (C_{p,t} - P_{p,t}(\omega))^{+},$$

where $(x)^{+} = \max(x,0)$.

From the distribution of $L_p(\omega)$ one may compute:

- expected loss;
- probability of shortfall;
- quantile-based risk measures;

- other performance indicators relevant to the structure.

These measures depend explicitly on structural parameters such as priority ordering and position size, making them suitable for comparative analysis across alternative configurations.

## 6.4 Structural thickness and risk allocation

A central design variable in many financial structures is the thickness or capacity of each position. Thickness determines how variability in inflows is absorbed and redistributed across the hierarchy of claims.

Let $\acute{N}_p$ denote the reference notional of position $p$. For a given stochastic inflow process, changes in $\acute{N}_p$ alter the distribution of losses and payments across all positions. The framework therefore allows the mapping

$$\acute{N}_p \rightarrow \text{distribution of } P_{p,t}(\omega)$$

to be analysed systematically.

This mapping provides a quantitative basis for evaluating how structural modifications shift risk and expected returns between senior and junior positions. It also supports sensitivity analysis with respect to structural parameters.

## 6.5 Structural evaluation and optimisation

Because all performance and risk metrics derive from scenario-dependent payment streams, structural evaluation can be formulated as a computational problem. For a given inflow model, alternative configurations of positions, trigger conditions, and allocation rules can be simulated and compared under a common set of scenarios.

Design of financial structures may therefore be cast as an optimisation problem:

$$\max_{\{\Theta_p\}} J\left(\{P_{p,t}(\omega)\}\right),$$

subject to structural, regulatory, or risk constraints. The objective functional $J$ may represent expected return, risk-adjusted performance, capital efficiency, or other design criteria.

The separation between stochastic inflow generation and deterministic allocation allows structural parameters to be varied while maintaining a consistent underlying scenario set. This facilitates systematic comparison of alternative configurations and transparent exploration of design trade-offs within a common stochastic environment. The resulting optimisation problems may nevertheless be high-dimensional and non-smooth when trigger-driven state transitions are present, and practical implementation may rely on restricted parametric families, heuristics, or robust search methods.

## 6.6 Limitations and scope

The proposed framework provides a formal and transparent representation of financial structures as computable allocation systems. Its scope and limitations follow from this level of abstraction and from the practical conditions under which structured finance models are implemented.

**Scenario complexity and computational scale.** The inflow-generation layer may induce a high-dimensional scenario space. Even with limited event types per unit, joint evolution across units and time can lead to combinatorial growth. In large transactions, tractable evaluation therefore relies on sampling, factor structures, clustering, or related approximation methods. The framework is compatible with such approaches but does not eliminate the intrinsic complexity of scenario generation.

**Granularity and model risk.** Representing inflows at the level of individual cash-flow generating units requires assumptions regarding event dynamics and dependence structures. In practice, data limitations and model uncertainty may influence the precision of structural evaluation. The framework clarifies how inflow assumptions propagate through allocation rules, but it does not replace the need for careful calibration and validation of the underlying stochastic model.

**Deterministic allocation and discretionary actions.** The allocation operator is deterministic conditional on realised inflows, the current structural state, and formally specified rules. This reflects the contractual logic of most waterfall mechanisms, yet some real-world arrangements include discretionary actions or soft triggers that are not fully codified. Such elements can be incorporated only to the extent that they can be expressed as explicit state variables or policy rules within the computational representation.

**Information frictions and operational discrepancies.** The framework assumes that allocation at time $t$ is determined by realised inflows and the current structure state. In practice, reporting delays, data revisions, or reconciliation procedures may generate differences between computed and realised payments. These frictions can be modelled as uncertainty in observed inflows or state variables but may limit real-time verification.

**Legal-to-computational translation.** Translating contractual documentation into a formally specified allocation operator and trigger set requires careful interpretation. While explicit computational representation enhances transparency, the mapping from legal language to executable rules must be validated through testing and scenario-based verification (see also (Scala 2025) for a legal–technical approach to contract-embedded executable waterfall logic).

**Feedback from structure to inflows.** The baseline framework separates inflow generation from allocation. This separation improves analytical clarity but may abstract from feedback channels through which structural features influence asset performance or incentives. Such interactions can be incorporated by allowing inflow dynamics to depend on structure-level state variables, extending the baseline representation.

**Optimisation tractability.** Structural design can be formulated as an optimisation problem, but the resulting design space may be high-dimensional and non-smooth. Trigger-driven regime changes can introduce discontinuities in outcome metrics, making global optimisation computationally demanding. In practice, structured search methods or restricted parametric families may provide more robust design approaches.

These considerations delimit the scope of the framework while indicating directions for extension and practical implementation within a computable architecture.

# 7. Comparative analysis under a common stochastic revenue environment

## 7.1 Purpose of the comparison

This section illustrates how the framework can be used to compare heterogeneous financial structures under a common stochastic inflow environment. The objective is not to calibrate a specific transaction, but to show how the same realised revenue paths can be passed through alternative allocation mechanisms and how differences in outcomes can then be attributed to contractual design.

We consider three stylised structures:

| Structure | Allocation logic | Reserve | Trigger | Risk sharing |
| --- | --- | --- | --- | --- |
| A | Sequential waterfall | No | No | Low |
| B | Sequential waterfall with reserve | Yes | Yes | Medium |
| C | Proportional revenue sharing | No | No | High |

All structures receive the same inflow sequence $F_t(\omega)$. They differ only in the allocation operator $A^{(k)}$, with $k \in \{A, B, C\}$.

## 7.2 Common stochastic inflow process

The inflow process is constructed from a generic revenue proxy rather than from a securitisation-specific asset pool. Let $X_t$ denote an observed activity or revenue index. The inflow entering the structure is

$$F_t(\omega) = \alpha X_t(\omega),$$

where $\alpha > 0$ is a scale parameter.

For the numerical illustration, the volatility scale is calibrated from the monthly log-returns of a traffic-based infrastructure proxy, inspired by the calendar- and seasonally-adjusted German truck-toll-mileage index. The empirical series is used only to calibrate the volatility of the inflow process, not to estimate a long-term trend or to conduct an empirical study of toll revenues.

Let

$$x_t = \log X_t.$$

Simulated scenarios are generated as a zero-drift lognormal random walk,

$$x_t(\omega) = x_{t-1}(\omega) + \varepsilon_t(\omega),$$

with

$$\varepsilon_t(\omega) \sim N(0, \sigma^2),$$

where $\sigma$ is the monthly volatility estimated from the observed proxy. The same simulated paths $F_{1:T}(\omega)$ are then passed through each allocation structure.

Baseline parameters are reported in Appendix C.

## 7.3 Structure A: sequential waterfall

Structure A is a standard sequential allocation rule. Let $D_t$ denote the senior debt service due at time $t$. Available funds are allocated first to the senior claim and then to the junior or residual claim.

$$P_{S,t}^{A}(\omega)=min\{F_t(\omega),D_t\},$$

$$P_{J,t}^{A}(\omega)=max\{F_t(\omega)-P_{S,t}^{A}(\omega),0\}.$$

The unpaid senior amount is

$$L_{S,t}^{A}(\omega)=D_t-P_{S,t}^{A}(\omega).$$

This structure concentrates downside risk on the junior position while protecting the senior position whenever inflows are sufficient to cover debt service.

## 7.4 Structure B: reserve account and trigger-based waterfall

Structure B extends the sequential waterfall by introducing a reserve account $R_t$ and a trigger rule.

Let $B$ denote the minimum reserve buffer and let $\acute{R}$ denote the reserve target level. The trigger indicator is

$$\chi_t(\omega)=1_{\{R_t(\omega)<B\}}.$$

Available resources before payments are

$$A_t(\omega)=F_t(\omega)+R_t(\omega).$$

Senior payments are

$$P_{S,t}^{B}(\omega)=min\{A_t(\omega),D_t\}.$$

Resources remaining after senior payment are

$$\widetilde{R}_{t+1}(\omega)=A_t(\omega)-P_{S,t}^{B}(\omega).$$

When the reserve is below target, residual funds are trapped and used to replenish the reserve. Junior payments are therefore

$$P_{J,t}^{B}(\omega)=\begin{cases} 0, & R_t(\omega)<B, \\ max\{\widetilde{R}_{t+1}(\omega)-\acute{R},0\}, & R_t(\omega)\geq B. \end{cases}$$

The reserve evolves according to

$$R_{t+1}(\omega)=\widetilde{R}_{t+1}(\omega)-P_{J,t}^{B}(\omega).$$

This structure sacrifices part of the junior upside in order to reduce liquidity stress and improve the stability of senior payments.

## 7.5 Structure C: proportional revenue-sharing

Structure C is not a debt waterfall. Instead, it represents a proportional participation in realised revenues.

Let $\lambda \in (0,1)$ denote the investor share. Investor payments are

$$P_{I,t}^{C}(\omega)=\lambda F_t(\omega),$$

while sponsor payments are

$$P_{O,t}^{C}(\omega)=(1-\lambda)F_t(\omega).$$

There is no reserve account, trigger mechanism, or priority ordering. Upside and downside are shared proportionally between participants.

## 7.6 Comparative metrics

For each structure $k \in \{A,B,C\}$ and each position $p$, the simulation generates a payment path

$$\{P_{p,t}^{k}(\omega)\}_{t=1}^{T}.$$

The total payment received by position $p$ is

$$P_p^k(\omega)=\sum_{t=1}^{T} P_{p,t}^{k}(\omega).$$

Expected payments are

$$E[P_p^k],$$

and payment volatility is measured by

$$Vol(P_p^k)=\sqrt{Var(P_p^k)}.$$

When promised payments are defined, losses are

$$L_p^k(\omega)=\left(\acute{P}_p^k - P_p^k(\omega)\right)^{+¿,¿}$$

where $\acute{P}_p^k$ denotes the benchmark payment.

This loss measure is defined only for positions with a promised or benchmark payment. For proportional revenue-sharing positions, where no promised payment is specified, the tail-loss metric is not applicable unless an external benchmark is introduced.

In the numerical illustration, tail loss is reported as the empirical 95th percentile of the cumulative loss distribution,

$$TL_{p,0.95}^k = q_{0.95}\left(\{L_p^k(\omega_m)\}_{m=1}^M\right).$$

It is therefore a loss quantile, not an expected shortfall.

For reserve-based structures, liquidity sustainability is measured by

$$Pr\left(\min_{t\le T} R_t(\omega) < B\right).$$

Condition (V4) is satisfied whenever

$$Pr\left(\min_{t\le T} R_t(\omega) < B\right) \le \eta.$$

## 7.7 Comparative results

The illustrative simulation uses $M=10,000$ scenarios and a horizon of $T=60$ periods. The monthly volatility of the inflow process is calibrated from the observed revenue proxy, while the drift is set to zero in order to avoid imposing a long-term growth or decline assumption. The same inflow scenarios are used for all three structures.

The resulting position-level metrics are reported below. The result table should not be read as ranking the three structures in general. It reports the consequences of one illustrative parameterisation. Its purpose is to show that expected payments, volatility, tail losses, and sustainability breaches can be computed consistently across distinct allocation mechanisms under the same stochastic inflow paths.

| Structure | Mean payment | Volatility | 95% tail loss | Reserve breach probability | V4 satisfied |
|---|---|---|---|---|---|
| A | 5485.146 | 193.301 | 562.328 | n.a. | n.a. |
| B | 5515.218 | 184.336 | 520.774 | 0.641 | no |

| Structure | Mean payment | Volatility | 95% tail loss | Reserve breach probability | V4 satisfied |
|---|---|---|---|---|---|
| C | 3977.934 | 268.283 | n.a. | n.a. | n.a. |

The same comparison can be visualised through the distribution of cumulative investor payments across scenarios. Figure 3 shows that the three allocation operators generate different payment distributions despite being evaluated under identical stochastic inflow paths.

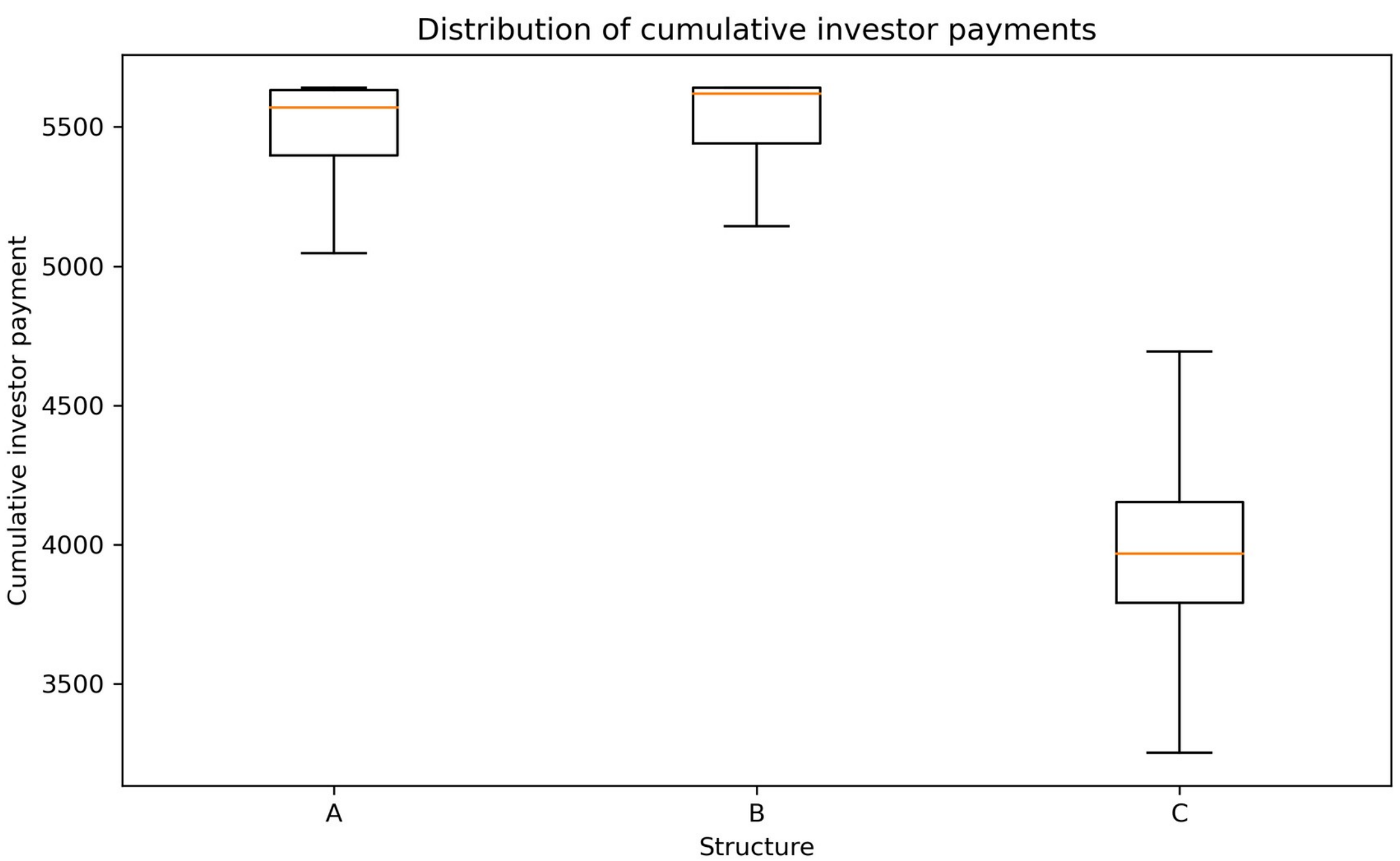

*Distribution of cumulative investor payments under the three allocation structures. The same inflow scenarios are passed through each allocation operator.*

## 7.8 Interpretation

The illustrative exercise yields three observations.

First, the reserve-and-trigger structure changes the senior payment distribution relative to the simple sequential waterfall. In the chosen parameterisation, Structure B delivers a higher mean senior payment than Structure A, with lower payment volatility and lower tail loss. This occurs because the reserve account absorbs part of the downside in low-inflow scenarios, reducing senior shortfalls that would otherwise occur under the simple waterfall.

Second, the same structure fails the operational sustainability condition (V4). The estimated reserve breach probability is $0.641$, above the tolerance level used in the illustration. This is not a calibration result for any real transaction. Rather, it demonstrates the role of the feasibility framework: a design may be structurally admissible and even improve senior payment metrics while still failing an operational sustainability constraint.

Third, the proportional revenue-sharing structure produces a different risk-return profile. Since Structure C has no priority ordering and no promised debt service, the tail-loss metric is not applicable. Its lower mean investor payment and higher volatility reflect the fact that the investor participates directly in realised revenues rather than receiving a priority-protected debt-like claim.

Taken together, these observations illustrate the operational value of separating inflow generation from allocation design. The inflow process defines the common stochastic environment; the allocation operator determines how that environment is transformed into payments, losses, and sustainability outcomes.

## 7.9 From comparison to design

The same architecture can also be used for constrained design. Let $d$ denote a vector of structural parameters, such as debt service levels, reserve targets, trigger thresholds, or revenue-sharing fractions. A generic design problem can be written as

$$\max_{d \in D} U(d),$$

subject to

$$Pr\left(\min_{t \le T} R_t^{(d)}(\omega) < B\right) \le \eta,$$

together with any participation, regulatory, or risk constraints.

The comparative exercise is therefore not only descriptive. It provides the computational basis for selecting among alternative structures or optimising design parameters under common stochastic conditions. In a full design exercise, candidate structures would be filtered by (V1)–(V4), not only by the sustainability condition (V4).

# 8. Relation to structured finance practice

## 8.1 Financial structures as operational systems

The framework developed in this paper corresponds closely to the operational logic of many structured finance arrangements. Securitisation vehicles, project finance structures, structured credit products, and similar arrangements are defined by payment waterfalls, priority rules, reserve mechanisms, and trigger conditions that determine how realised inflows are applied across positions.

The allocation operator introduced above provides a direct computational representation of this logic while separating allocation from the stochastic modelling of underlying cash-flow generation.

## 8.2 Transparency and comparability

Representing financial structures through explicit allocation operators and stochastic inflow processes provides a basis for systematic comparison across alternative configurations.

This transparency is difficult to obtain when structures are analysed through ad hoc deal-specific models or proprietary implementations. By expressing inflow generation and allocation rules within a unified computational framework, the approach supports consistent evaluation and documentation of structural features across transactions and asset classes.

## 8.3 Computational implementation

The separation between stochastic inflow generation and deterministic allocation enables modular implementation. Scenario generation for underlying assets can be developed independently of the allocation mechanism, while alternative structural configurations can be evaluated using the same simulated inflow paths. This modularity supports iterative design, sensitivity analysis, validation, and auditability.

Because the allocation operator is deterministic, realised payment streams are reproducible given a specified inflow path, structural state, and set of contractual parameters. This property is useful for communication among originators, investors, rating agencies, regulators, and other stakeholders.

## 8.4 Relation to existing structuring methodologies

Detailed methodologies for modelling securitised asset pools and constructing payment waterfalls have been developed in industry practice. These approaches typically integrate asset-level cash-flow modelling, scenario construction, and structural analysis within transaction-specific workflows.

The framework introduced here abstracts from deal-specific implementations and isolates the computational elements common to them. By separating stochastic inflow generation, deterministic allocation, and structural design, it provides a general representation applicable across a broad class of financial structures while remaining compatible with existing modelling practices. This explicit representation also provides a foundation for executable and verifiable implementations of financial structures, as explored in recent work on contract-embedded waterfall logic (Scala 2025).

## 8.5 Implications for structural design

Viewing financial structures as structured allocation systems highlights the role of contractual design in shaping the distribution of risk and return. Structural parameters such as position size, priority ordering, maturity profile, reserve policy, and trigger thresholds can be treated as explicit variables within a computational design problem.

Although motivated by structured finance, the representation also applies to revenue-based arrangements in which stochastic resources are distributed across contractual claims. The proportional revenue-sharing example in Section 7 illustrates this broader applicability without requiring a securitisation-specific inflow model.

The explicit algorithmic representation of allocation rules enhances inspectability. Investors can analyse how alternative scenarios affect payments across positions, while regulators and other stakeholders can examine the consistency and robustness of contractual mechanisms. Under the proposed representation, uncertainty affects payment outcomes through stochastic inflows and state transitions, while the allocation mechanism itself remains a deterministic mapping conditional on the realised state and contractual parameters.

# 9. Conclusion

This paper has formalised financial structures as structured allocation systems that map stochastic inflows into ordered payments across claims. Its main contribution is to show that structurally distinct contractual mechanisms can be represented as allocation operators and evaluated under identical stochastic inflow paths.

The framework is representational rather than a new pricing or asset-side risk model. Its role is to make the allocation logic explicit, so that contractual mechanisms can be simulated, inspected, and compared under common stochastic conditions.

The framework addresses a gap that is largely independent of any specific asset class or market segment. Rather than proposing another cash-flow model, it provides a general computational representation in which financial structures themselves can be treated as explicit, comparable, and verifiable objects. The key separation is between the stochastic generation of inflows and the deterministic allocation mechanisms that transform those inflows into position-level payment streams. It distinguishes five conceptually separate layers: economic activities, financeable representations, admissible structures, feasible designs, and sustainable designs. This hierarchy clarifies where economic assumptions enter the model and where contractual design operates.

The comparative exercise illustrates this scope in operational terms. A sequential waterfall, a reserve-and-trigger waterfall, and a proportional revenue-sharing arrangement are represented within the same operator language and evaluated under the same stochastic inflow paths. This shows how the framework can compare structurally distinct contractual mechanisms without changing the underlying stochastic environment.

The framework applies most directly to arrangements in which allocation rules and trigger conditions are explicitly specified and mechanically enforceable. Discretionary actions, renegotiation, reporting frictions, and feedback from structural design to inflow generation may require extensions of the state space or additional policy rules. These extensions do not alter the basic architecture, but they mark important directions for further work.

More generally, the framework separates economic activity, financial representation, structural admissibility, economic feasibility, and operational sustainability into distinct analytical layers. This separation allows financial structures to be studied as computable economic mechanisms whose design, behaviour, and resilience can be analysed within a common stochastic environment.

# Statements and Declarations

The authors received no funding; no competing interests present.

**Author Contributions** All authors contributed equally to the paper.

# A: Formal specification of the allocation operator

This appendix provides a formal specification of the allocation framework introduced in the main text and states the fundamental properties satisfied by admissible allocation operators. The objective is to make explicit the state-dependent deterministic structure underlying the computational representation.

## Structural state and allocation operator

Let time be discrete, $t = 0, 1, \ldots, T$. Let $(\Omega, F, P)$ be a probability space supporting a non-negative inflow process $\{F_t(\omega)\}_{t=0}^{T}$.

A financial structure is characterised by a structural state vector $S_t(\omega)$ taking values in a state space $S$. The state encodes all contractual variables required to determine admissible payments and their evolution over time. Depending on the application, $S_t$ may include residual or reserve balances, outstanding tranche balances, accrued but unpaid amounts, trigger indicators, and cumulative performance measures.

The allocation mechanism is represented by a deterministic state-transition operator

$$(P_{1:P,t}(\omega), S_{t+1}(\omega)) = A_t\left(F_t(\omega), S_t(\omega)\right),$$

where $P_{p,t}(\omega) \geq 0$ denotes the payment made to position $p \in \{1, \ldots, P\}$ at time $t$.

Thus, conditional on realised inflows and current state, both the payment vector and next-period state are uniquely determined. Stochasticity in payments arises solely from stochastic inflows or stochastic state transitions induced by those inflows.

In many implementations, all relevant historical information is encoded in the current state $S_t$, so that the operator admits a Markov representation

$$A_t : R_{+¿ \times S \to R_{+¿^{P} \times S.¿}¿}$$

## Residual cash and conservation of funds

Let $R_t(\omega)$ denote liquid funds carried into period $t$, included as a component of the state vector $S_t(\omega)$. Define total available funds at time $t$ as

$$A_t(\omega)=F_t(\omega)+R_t(\omega).$$

An admissible allocation operator must satisfy conservation of funds:

$$\sum_{p=1}^{P} P_{p,t}(\omega)+R_{t+1}(\omega)=F_t(\omega)+R_t(\omega),$$

with $R_t(\omega)\geq 0$ for all $t,\omega$.

## Priority structure

Positions are indexed in strict priority order: if $p<q$, then position $p$ has higher priority than position $q$.

An allocation operator is priority-consistent if, for each $t$ and $\omega$, payments to a junior position can occur only after all admissible payments to higher-priority positions have been satisfied according to the contractual rules encoded in the operator.

## Fundamental properties of sequential allocation operators

Sequential waterfall structures form a central subclass of admissible operators. Such operators satisfy the following properties.

**Non-negativity.**

$$P_{p,t}(\omega)\geq 0\ \forall\ p,t,\omega.$$

**Conservation.** For all $t$ and $\omega$,

$$\sum_{p=1}^{P} P_{p,t}(\omega)+R_{t+1}(\omega)=F_t(\omega)+R_t(\omega).$$

**Priority consistency.** If $p<q$, then admissible payments to position $q$ occur only after admissible payments to positions $1,\ldots,p$ have been satisfied.

**Monotonicity within a fixed sequential regime.** Let $\widetilde{F}_t(\omega)\geq F_t(\omega)$ for all $t$, and apply the same fixed-regime sequential waterfall to the two inflow paths while holding the initial state fixed. Then payments to more senior positions cannot decrease:

$$\widetilde{P}_{p,t}(\omega) \geq P_{p,t}(\omega)$$

for all $p$ until contractual caps for that position become binding.

This property applies to fixed sequential regimes. It need not hold for individual positions when the increase in inflows changes trigger status, activates reserve replenishment, or switches the allocation regime. In such cases, additional resources may be redirected to state-dependent contractual objectives rather than paid immediately to a given position.

*Proof sketch.* Within a fixed sequential regime, allocation applies available funds in priority order. Increasing available funds weakly increases the funds available to each senior position before lower-priority positions are considered; hence senior payments cannot decrease unless the contractual cap is already met. The caveat follows because trigger-driven operators may change the mapping itself when the state crosses a contractual threshold. ▫

**Existence and uniqueness.** For any admissible inflow path $\{F_t(\omega)\}$ and initial state $S_0(\omega)$, a sequential allocation operator produces a unique sequence of payments and states, because at each period the output is obtained by a finite deterministic procedure.

# B: Comparative metrics and Monte Carlo estimators

This appendix defines the metrics used in the comparative exercise of Section 7 and gives the corresponding Monte Carlo estimators. The notation is deliberately structure-neutral: the same estimators apply to the sequential waterfall, the reserve-and-trigger waterfall, and the proportional revenue-sharing structure.

## Scenario-level payment paths

Let $k \in \{A, B, C\}$ index the structure and let $p$ index a position within that structure. For scenario $\omega_m$, the allocation operator produces the payment path

$$\{P^k_{p,t}(\omega_m)\}_{t=1}^{T}.$$

The cumulative payment to position $p$ is

$$\Pi^k_p(\omega_m) = \sum_{t=1}^{T} P^k_{p,t}(\omega_m).$$

When a benchmark or promised payment is defined, denote it by

$$\acute{\Pi}_p^k.$$

For debt-like positions, $\acute{\Pi}_p^k$ may represent the cumulative contractual debt service. For proportional revenue-sharing positions, it may be omitted or replaced by a target payment chosen by the analyst.

For debt-like positions, $\acute{\Pi}_p^k$ represents the cumulative contractual debt service. In the illustrative structures A and B, the benchmark for the senior position is therefore

$$\acute{\Pi}_S^k = \sum_{t=1}^{T} D_t, k \in \{A, B\}.$$

For proportional revenue-sharing positions, no promised payment is specified; $\acute{\Pi}_p^k$ is therefore omitted unless the analyst introduces an external target or benchmark.

## Loss and shortfall variables

When a benchmark payment is defined, the cumulative shortfall of position $p$ in structure $k$ is

$$L_p^k(\omega_m) = \left(\acute{\Pi}_p^k - \Pi_p^k(\omega_m)\right)^{+\text{¿},\text{¿}}$$

where $(x)^{+\text{¿}=max\{x,0\}\text{¿}}$.

The corresponding loss rate is

$$\lambda_p^k(\omega_m) = \frac{L_p^k(\omega_m)}{\acute{\Pi}_p^k}, \acute{\Pi}_p^k > 0.$$

The impairment indicator is

$$I_p^k(\omega_m) = 1_{\{\Pi_p^k(\omega_m) < \acute{\Pi}_p^k\}}.$$

These quantities are computed only for positions for which a contractual or benchmark payment is meaningful.

## Liquidity-sustainability variable

For structures with reserve or liquidity accounts, define

$$Q^k(\omega_m) = 1_{\left\{\min_{t \le T} R_t^k(\omega_m) < B\right\}}.$$

Thus, $Q^k(\omega_m)=1$ if structure $k$ breaches the liquidity buffer in scenario $\omega_m$ and $Q^k(\omega_m)=0$ otherwise.

For structures without a reserve account, this metric is not applicable unless an equivalent state variable is introduced.

## Monte Carlo estimators

Let $M$ be the number of simulated or bootstrapped scenarios.

The expected cumulative payment is estimated by

$$\hat{E}[\Pi_p^k]=\frac{1}{M}\sum_{m=1}^{M}\Pi_p^k(\omega_m).$$

Payment volatility is estimated by

$$\widehat{Vol}(\Pi_p^k)=\left[\frac{1}{M-1}\sum_{m=1}^{M}\left(\Pi_p^k(\omega_m)-\hat{E}[\Pi_p^k]\right)^2\right]^{1/2}.$$

When a benchmark payment is defined, expected loss is estimated by

$$\hat{E}[L_p^k]=\frac{1}{M}\sum_{m=1}^{M}L_p^k(\omega_m),$$

and the impairment probability by

$$\hat{\psi}_p^k=\frac{1}{M}\sum_{m=1}^{M}I_p^k(\omega_m).$$

A tail-loss statistic at confidence level $a\in(0,1)$ can be computed as the empirical quantile

$$\widehat{TL}_{p,a}^k=\hat{q}_a\left(\{L_p^k(\omega_m)\}_{m=1}^{M}\right).$$

In Section 7, we use $a=0.95$.

For reserve-based structures, the liquidity-buffer breach probability is estimated by

$$\hat{q}_B^k=\frac{1}{M}\sum_{m=1}^{M}Q^k(\omega_m).$$

Condition (V4) is satisfied empirically when

$$\hat{q}_B^k\leq\eta.$$

## Comparative use

The estimators above are applied to the same set of inflow paths for all structures. This common-random-environment design reduces noise in pairwise comparisons and ensures that differences in estimated metrics arise from the allocation operators rather than from different stochastic samples.

For two structures $k$ and $\ell$, differences in expected cumulative payment can be estimated as

$$\hat{\Delta}_p^{k,\ell} = \frac{1}{M} \sum_{m=1}^{M} \left[ \Pi_p^k(\omega_m) - \Pi_p^\ell(\omega_m) \right],$$

whenever the positions being compared have a consistent economic interpretation.

Analogous differences may be computed for loss, volatility, tail loss, or liquidity-buffer breach probabilities. These pairwise comparisons are the numerical counterpart of the allocation-operator representation introduced in the main text.

# C: Common revenue environment and scenario generation

This appendix documents the stochastic inflow environment used for the comparative illustration in Section 7. The purpose of the specification is not to provide an asset-side model, but to generate a common set of revenue paths that can be passed through alternative allocation structures.

## Revenue proxy

Let $\{X_t\}_{t=1}^T$ denote an observed activity or revenue proxy. In the numerical illustration, $X_t$ is inspired by a traffic-based infrastructure index. The empirical series is used only to calibrate the volatility scale of the inflow process.

The inflow entering the financial structure is defined as

$$F_t(\omega) = \alpha X_t(\omega),$$

where $\alpha > 0$ is a scale parameter. The same scaling is applied across all structures.

## Volatility calibration

Let

$$x_t = \log X_t .$$

The monthly log-return is

$$r_t = x_t - x_{t-1} .$$

The volatility parameter used in the simulation is the sample standard deviation of monthly log-returns,

$$\hat{\sigma} = \left[ \frac{1}{n-1} \sum_{t=2}^{n} \left( r_t - \acute{r} \right)^2 \right]^{1/2} ,$$

where

$$\acute{r} = \frac{1}{n-1} \sum_{t=2}^{n} r_t .$$

In the numerical illustration, the estimated monthly volatility is

$$\hat{\sigma} = 0.0148 .$$

The empirical mean return is not used as a drift term. The drift is set to zero deliberately, so that the observed proxy calibrates the scale of fluctuations but does not impose a long-term growth or decline assumption.

## Scenario generation

Starting from the last observed index level $X_0$, simulated log-index paths are generated as

$$x_t^{(m)} = x_{t-1}^{(m)} + \varepsilon_t^{(m)} ,$$

with

$$\varepsilon_t^{(m)} \sim N\left(0, \hat{\sigma}^2\right) ,$$

for scenarios $m = 1, \ldots, M$ and periods $t = 1, \ldots, T$.

The corresponding index and inflow paths are

$$X_t^{(m)} = \exp\left(x_t^{(m)}\right) ,$$

and

$$F_t^{(m)} = \alpha X_t^{(m)} .$$

The collection

$$\left\{ F_{1:T}^{(m)} \right\}_{m=1}^{M}$$

is generated once and then reused across all structures considered in Section 7.

## Common random environment

For each scenario $m$, the same inflow path is passed through:

1. the sequential waterfall;
2. the reserve-and-trigger waterfall;
3. the proportional revenue-sharing structure.

This common-random-environment design ensures that differences in payment distributions, losses, reserve dynamics, and sustainability breaches are attributable to allocation rules rather than to different stochastic inflow samples.

## Baseline parameters

The numerical illustration uses the following baseline parameters:

| Parameter | Symbol | Value |
|---|---:|---:|
| Number of scenarios | $M$ | $10{,}000$ |
| Horizon | $T$ | 60 |
| Revenue scaling | $\alpha$ | 1.0 |
| Initial index level | $X_0$ | 94.3 |
| Monthly log-return volatility | $\hat{\sigma}$ | 0.0148 |
| Senior debt service | $D_t$ | 94.0 |
| Minimum reserve buffer | $B$ | 20.0 |
| Reserve target | $\acute{R}$ | 40.0 |
| Initial reserve | $R_0$ | 40.0 |
| Investor revenue share | $\lambda$ | 0.70 |
| Sustainability tolerance | $\eta$ | 0.05 |

# D: Design-space exploration

The comparative framework of Section 7 can be extended from fixed-structure evaluation to design-space exploration. Rather than evaluating a single allocation rule, the analyst may parameterise the allocation operator and compare admissible designs under the same stochastic inflow environment.

## Parameterised allocation operators

Let

$$d \in D$$

denote a vector of structural parameters. Depending on the application, $d$ may include debt-service schedules, reserve targets, trigger thresholds, revenue-sharing fractions, priority relations, contractual caps, or maturity profiles.

For each admissible parameter vector $d$, the allocation rule becomes

$$A^{(d)}.$$

Applying this operator to scenario $\omega$ produces payment paths and state trajectories,

$$\left( \{ P_{p,t}^{(d)}(\omega) \}_{p,t}, \{ S_t^{(d)}(\omega) \}_t \right).$$

## Feasibility filtering

Not every admissible allocation rule is economically feasible or operationally sustainable. Let

$$F \subseteq D$$

denote the subset of designs satisfying the feasibility and sustainability conditions introduced in Section 5.6:

$$F = \{ d \in D : \text{conditions \textbackslash(V1\textbackslash)–\textbackslash(V4\textbackslash) are satisfied} \}.$$

For example, a design may be required to satisfy investor acceptability constraints,

$$\psi_p(d) \le \varepsilon_p,$$

originator participation constraints,

$$\sum_{p \in O} E[\Pi_p^{(d)}] - \kappa \ge \rho_0,$$

and liquidity-sustainability constraints,

$$Pr\left(\min_{t \le T} R_t^{(d)}(\omega) < B\right) \le \eta .$$

The design problem is therefore posed over $F$, not over the full parameter space $D$
.

## Objective functions

A structural design objective can be written abstractly as

$$U(d) = U\left(\{P_{p,t}^{(d)}(\omega)\}, \{S_t^{(d)}(\omega)\}\right),$$

where $U$ may encode expected payments, volatility, loss measures, liquidity breaches, or stakeholder-specific performance criteria.

The corresponding constrained design problem is

$$d^{\star} = \arg\max_{d \in F} U(d).$$

The framework does not prescribe a unique objective function. Its role is to provide a common computational representation in which different objectives and constraints can be evaluated consistently.

## Interpretation

Design-space exploration is a direct consequence of the allocation-operator representation. Once financial structures are represented as explicit mappings from inflows and state variables to payments and updated states, alternative designs can be compared under identical stochastic environments.

This does not imply that the framework solves the economic problem of optimal security design in general. Pricing, preferences, market-clearing, and strategic behaviour remain external to the baseline representation unless explicitly added. The contribution is narrower: it provides a tractable computational setting in which contractual allocation mechanisms can be specified, filtered by feasibility conditions, and compared or optimised with respect to declared performance criteria.